\documentclass[12pt]{article}

\usepackage{amsmath,graphicx}
\usepackage{simplewick}
\usepackage{epsf}
\usepackage{graphicx,epsfig}
\usepackage{amsfonts}
\usepackage{amssymb}
\usepackage[nosort]{cite}
\usepackage{setspace}

\def\bk{{\bf k}}

\def\bx{{\bf x}}

\def\CH{{\cal H}}
\def\CL{{\cal L}}
\def\CO{{\cal O}}

\def\mpl{M_{\rm P}}
\def\half{\frac{1}{2}}

\makeatletter
\renewcommand\section{\@startsection {section}{1}{\z@}%
                                 {-3.5ex \@plus -1ex \@minus -.2ex}%
                                   {2.3ex \@plus.2ex}%
                                   {\normalfont\large\bfseries}}
\renewcommand\subsection{\@startsection{subsection}{2}{\z@}%
                                   {-3.25ex\@plus -1ex \@minus -.2ex}%
                                     {1.5ex \@plus .2ex}%
                                     {\normalfont\bfseries}}
\renewcommand\subsubsection{\@startsection{subsubsection}{3}{\z@}%
                                   {-3.25ex\@plus -1ex \@minus -.2ex}%
                                     {1.5ex \@plus .2ex}%
                                     {\normalfont\itshape}}
\makeatother

\newcommand{\Letter}{
\setlength{\textwidth}{16.5cm}
   \setlength{\textheight}{22.6cm}
    \hoffset=-0.5in
\voffset=-2.1cm }

\Letter

\begin{document}
\newcommand{\be}{\begin{equation}}
\newcommand{\ee}{\end{equation}}
\newcommand{\bea}{\begin{eqnarray}}
\newcommand{\eea}{\end{eqnarray}}
\newcommand{\barr}{\begin{array}}
\newcommand{\earr}{\end{array}}

\thispagestyle{empty}
\begin{flushright}
\end{flushright}

\vspace*{0.3in}
\begin{spacing}{1.1}

\begin{center}
{\large \bf Primordial Features as Evidence for Inflation}

\vspace*{0.5in} {Xingang Chen}
\\[.3in]
{\em Center for Theoretical Cosmology, \\
Department of Applied Mathematics and Theoretical Physics, \\
University of Cambridge, Cambridge CB3 0WA, UK} \\[0.3in]
\end{center}

\begin{center}
{\bf
Abstract}
\end{center}
\noindent
In the primordial universe, fields with mass much larger than the mass-scale of the event-horizon (such as the Hubble parameter in inflation) exist ubiquitously, and can be excited from time to time and oscillate quickly around their minima. These excitations can induce specific patterns in density perturbations, which record the time dependence of the scale factor of the primordial universe, thus provide direct evidence for the inflation paradigm or its alternatives. Such effects are conventionally averaged out in theoretical and data analyses, but can be accessible for experiments targeting on density perturbations with high multipoles.

\vfill

\newpage
\setcounter{page}{1}


\newpage

\section{Introduction}
\setcounter{equation}{0}

The inflation \cite{Guth:1980zm,Linde:1981mu,Albrecht:1982wi}, as the leading candidate paradigm for the primordial universe, has received strong support from observations of cosmic microwave background (CMB) and large scale structure (LSS) \cite{Komatsu:2010fb}. The simplest inflationary scenario not only explains the homogeneity and isotropy of the Universe, but also predicts that the density perturbations seeding the large scale structure are generated at superhorizon scales, and are approximately scale-invariant, Gaussian and adiabatic, all of which are verified by experiments to some extent.

Nonetheless, based on current observations, ambiguities and degeneracies still exit in terms of model-building. The specific inflation models remain illusive; in addition, there may be alternatives to inflation that have the same consequences on observables that we have been able to measure so far. To make further progress, there are at least two important questions.
The first is how to find evidence that would unambiguously distinguish inflation from its alternatives.
The second question is, if inflation is the correct paradigm, how to predict and measure new observables that will pin down the microscopic details. Similar questions apply to the alternative paradigms.

Because all viable models have to satisfy the current observations that the two-point correlation function (power spectrum) of the density perturbations is approximately scale-invariant, for the purpose of the second question, studying small deviations from the scale-invariance and measurable higher-point correlation functions (non-Gaussianities) become very important. For example, for inflation models, properties of primordial non-Gaussianities can be classified \cite{Chen:2010xk}, and if measurable, provide evidence for the interaction terms in the Lagrangian. Similar classification may be established for each of the alternative paradigms.

However the first question still remains unanswered from this line of research. For example, assuming single field models and imposing the condition that the power spectrum is approximately scale invariant, the consequences on non-Gaussianities can be systematically worked out along the line of \cite{Maldacena:2002vr,Seery:2005wm,Chen:2006nt,Khoury:2008wj,Noller:2011hd}, for either inflation or alternatives. The inflation models still predict approximately scale-invariant non-Gaussianities, while the attractor alternative paradigms \cite{ArmendarizPicon:2003ht,Khoury:2009my,Linde:2009mc,Khoury:2010gw,Baumann:2011dt} predict non-scale-invariant ones. However, as soon as we step away from this subset and consider multifield models, this sharp distinction will be lost. For example, for inflation models, multiple fields introduce various isocurvature modes that can have scale-dependent couplings to the curvature mode. Non-Gaussianities can be easily made non-scale-invariant if they are transferred from the isocurvature modes, for instance, when inflaton makes a non-constant turn in its trajectory in models of the type \cite{Chen:2009we,Chen:2009zp}. Reversely, scale-invariant non-Gaussianities may be achieved in multifield or non-attractor single field non-inflationary models \cite{Cai:2009fn,Brandenberger:2011gk}. In short, given general inflation scenarios, there is no generic prediction on how these non-Gaussianities should depend on scales.

So far the primordial tensor mode is regarded as the only possible solution regarding to the first question. The tensor modes from inflation models are approximately scale-invariant with a red tilt; for some models they are observable. Typical alternatives such as the cyclic model \cite{Khoury:2001wf,Khoury:2003rt} or string gas cosmology \cite{Brandenberger:1988aj,Brandenberger:2006xi} predict either non-observable tensor fluctuations or observable ones with blue-tilt. However, there are some important caveats for the tensor modes to achieve the goal unambiguously. Firstly, if we consider more general alternatives, scale-invariant and observable tensor modes are possible. The equation of motion obeyed by each polarization component of the tensor modes is the same as that by the massless scalar. So the tensor modes can be scale-invariant even in non-inflationary spacetime, just as the scalar. For it to be observable, we only need a large Hubble parameter. Scenarios of matter contraction \cite{Wands:1998yp,Finelli:2001sr} with large Hubble parameter are such explicit examples. Secondly, even for inflation, tensor modes are not guaranteed to be observable. While the best sensitivity for the tensor-to-scalar ratio achievable by experiments in the near future is $\Delta r \sim \CO(10^{-3})$, the inflation models predict anywhere between $r \sim \CO(10^{-1})$, for large field models, and $r\sim\CO(10^{-55})$, for small field models with TeV-scale reheating energy.

So it is very important to search for complimentary properties in the density perturbations that can serve as a model-independent general distinguisher between inflation and alternatives. This is the main purpose of this paper.

Before proceed, we would like to make a comment on the types of models we investigate. Arguably, inflation remains as the best available paradigm for the primordial universe. Its generic predictions naturally fit the data and its microscopic origin in term of fundamental theory is promising. Nonetheless, such opinions may be subject to personal taste; they are model-dependent and may even evolve with time. A more uncontroversial standard will be in terms of experimental data, and to ask what we can learn given the data by reverse engineering. So in this paper we will not discuss the important UV completion and model building aspects of the non-inflationary backgrounds. For the same reason, we will also not discuss which alternative of inflation is more natural than the others, for example, between expansion and contraction, attractor and non-attractor scenario.

\section{Bunch-Davies vacuum and resonance mechanism}
\label{Sec:BDandres}
\setcounter{equation}{0}

In nearly all models of primordial universe, the quantum fluctuations start their life in a vacuum that is mostly Bunch-Davies (BD). These fluctuations later exit the event horizon and become the seeds for the large scale structure. This applies to both inflationary and non-inflationary scenario, expansion and contraction universe, attractor and non-attractor evolution, single field and multifield model, curvaton and isocurvaton modes.

For example, consider the fluctuations of an effectively massless scalar field, $\delta\phi(\bx,t)$, in a general time-dependent background with scale factor $a(t)$,
\bea
L= \int d^3x  \left[ \frac{a^3}{2} (\dot {\delta\phi})^2 - \frac{a}{2} (\partial_i \delta\phi)^2 \right] ~.
\label{Ldeltaphi}
\eea
The conformal time $\tau$ is defined as $d\tau= dt/a$, and we will use dot to denote the derivative with respective to $t$ and prime to $\tau$. The event horizon\footnote{Here the event horizon is defined to be the maximum distance at $t$ by which two points are separated but can still communicate with each other from $t$ to $t_{\rm end}$. So it is $a(t)\int_t^{t_{\rm end}} dt/a = -a\tau$, where $\tau_{\rm end}$ is set to $0$.} in physical coordinates is therefore $|a\tau|$. A quantum fluctuation with comoving momentum $k$ is within the event horizon if $k>1/|\tau|$. In this limit, the equation of motion for the fluctuations approaches that in the Minkowski spacetime limit. Along with the quantization condition,
\bea
a^3 \delta\phi \dot{\delta\phi}^* - {\rm c.c.} = i ~,
\eea
the mode function in the subhorizon limit becomes
\bea
\delta\phi \to \frac{1}{a \sqrt{2k}} e^{-ik\tau} ~.
\label{BDphi}
\eea
We have chosen the positive-energy mode, which corresponds to the ground state of the Minkowski spacetime, to be the BD vacuum. The effect of the background time-dependence is incorporated adiabatically in (\ref{BDphi}). The most important and universal property of (\ref{BDphi}) is the oscillatory factor $e^{-ik\tau}$. Various prefactors depend on whether the form of Lagrangian (\ref{Ldeltaphi}) is canonical.

To give explicit examples of the time-dependent backgrounds, we take the scale factor to be of the general power-law,
\bea
a(t) = a(t_0) (t/t_0)^p ~.
\label{powerlaw}
\eea
Because we require that the quantum fluctuations exit the event horizon,
for $p>1$ we need an expansion phase, so $t$ runs from $0$ to $+\infty$; for $0<p<1$ we need a contraction phase, so $t$ runs from $-\infty$ to $0$; for $p<0$, we again need an expansion phase, so $t$ runs from $-\infty$ to $0$.
The conformal time $\tau$ is related to $t$ by $a\tau=t/(1-p)$, and $\tau$ always runs from $-\infty$ to $0$.
For example, $p>1$ corresponds to the inflation \cite{Guth:1980zm,Linde:1981mu,Albrecht:1982wi}, $p=2/3$ the matter contraction \cite{Wands:1998yp,Finelli:2001sr}, $p=1/3$ the pre-big-bang \cite{Gasperini:1992em,Enqvist:2001zp}, $0<p\ll 1$ the ekpyrotic (slowly contracting) phase \cite{Khoury:2001wf}, and $-1 \ll p <0$ the slowly expanding phase \cite{Piao:2003ty}.

To directly probe the universal BD vacuum, we need a high energy probe with wavelength much shorter than the event horizon. This can be achieved by introducing a small but highly oscillatory component in the background evolution \cite{Chen:2008wn}. Such a component resonates with the vacuum component which has the same physical wavelength. Because the BD vacuum has time-dependence, different momentum modes get resonated at different time. This effect is formulated in terms of the following integral,
\bea
\int d\tau B(t) e^{-iK\tau} + {\rm c.c.} ~.
\label{ResIntegral}
\eea
The factor $e^{-iK\tau}$ in the integrand is the universal BD oscillatory component, and $K$ is some comoving momentum. The factor $B(t)$ denotes the high energy probe mode we introduce. The integrand resonates when the two factors have the same frequency. Different momentum modes resonate one by one with the background, and in the mean while the phase of the background repeats due to oscillation. If we regard the repeated oscillation in $B(t)$ as a clock, the time-dependence of the scale factor is translated into the $k$-dependence of the integral through the resonance mechanism.
For example if we take $B(t)$ to be a periodic clock with frequency $\omega$, $B(t) \sim e^{i\omega t}$, (\ref{ResIntegral}) becomes proportional to\footnote{Using $\int_{-\infty}^\infty dx~ e^{if(x)} \approx - e^{\mp i3\pi/4} \sqrt{2\pi} e^{if_*}/\sqrt{\pm f''_*}$, for positive/negative $f''_*$, where the subscript ``${}_*$" denotes the resonant point $f'(x_*)=0$. Away from this point, the integrand $e^{i f(x)}$ is oscillating rapidly.}
\bea
\sim \sin \left[ \frac{p^2}{p-1} \frac{\omega}{H_0} \left( \frac{K}{k_r} \right)^{1/p} + {\rm phase}\right] ~,
\label{ResPowerlaw}
\eea
where the ``phase" denotes a $K$-independent constant, $k_r$ is the mode that resonates at $t_0$, and $H_0$ is the Hubble parameter at $t_0$. As we can see, the time-dependence of the scale factor is encoded inside the square bracket of (\ref{ResPowerlaw}) as a function of $K$-modes. For power-law background and periodic resonance, this function is the inverse function of the scale factor. If we take the exponential inflation limit $p\gg 1$ and study a range of modes $\Delta K$ satisfying $\ln (\Delta K/k_r) \ll p$, we have $(K/k_r)^{1/p} \to 1+ (1/p) \ln(K/k_r)$. So (\ref{ResPowerlaw}) goes to
\bea
\sin \left[ \frac{\omega}{H} \ln \frac{K}{k_r} + {\rm phase} \right] ~.
\label{CELform}
\eea
This is the leading resonance form found by Chen, Easther and Lim (CEL) for inflation \cite{Chen:2008wn}. As we can see, the CEL form is a special limit of the general resonance forms. In retrospect, the reason the argument of the sinusoidal function is proportional to $\ln K$ is that this is the inverse function of the exponential function in inflationary scale factor.

The distinctive oscillatory running behavior in the above resonant forms will not be changed by curvaton-isocurvaton couplings in multifield evolution. Any effect that also oscillates faster than the horizon time-scale generates additional resonance forms that superimpose onto each other. Any effect that varies much slower can only change the overall envelop of the resonance forms, by either changing their overall sizes or introducing scale dependent modulations.
This latter modulation can also be informative as we will see in more details later. But similar to the scale-dependence of non-oscillatory correlation functions that we mentioned in Introduction, these scale-dependence can be rather arbitrary in multifield models, so much less robust than the resonant running.

However, in terms of reverse engineering, we should also consider the possibility of non-periodic background oscillation components. A non-periodic background oscillation may cause resonance in a non-inflationary background, and conspire to give the same CEL form. For example, for arbitrary power law behavior (\ref{powerlaw}), we may engineer a background oscillation component to be of the form $B(t) \sim e^{i g \ln(t/t_0)}$, so that the resulting resonance behavior is
\bea
\sin \left[ \frac{g}{p-1} \ln \frac{K}{k_r} + {\rm phase} \right] ~,
\eea
which is the same as (\ref{CELform}). Therefore it becomes very important to search for standard clocks in physical systems. Such a clock should generate repeated perturbations with known time dependence, although not necessarily periodic. They should also be associated with a set of specific patterns that can be identified in observations.

\section{Spectator massive fields as standard clock}
\setcounter{equation}{0}

Massive fields with mass much larger than the horizon mass-scale $1/|a\tau|$ exist ubiquitously in models of primordial universe.\footnote{For $|p|\gg 1$, $1/|a\tau| \approx |H|$; for $|p|\ll 1$, $1/|a\tau| \gg |H|$.} Even when we think of single field models, in a UV completed context, what we have in mind is really models with many massive modes. The single field model is obtained as the low energy limit where the energy scale is comparable to or smaller than $1/|a\tau|$, after these massive modes are integrated out. This is a good approximation even if the massive modes get excited classically and oscillate around its minimum. But for our purpose, these oscillations are a good candidate for the physical clock that we are looking for. So let us look at more details of the classical behavior of a massive particle $\sigma$ in the power-law background.

The equation of motion is
\bea
\ddot \sigma + 3H\dot\sigma + m_\sigma^2 \sigma =0 ~,
\eea
where the Hubble parameter $H=p/t$. The solution is given in terms of Bessel functions. The asymptotic behavior of these Bessel functions at the limit $m_\sigma t\gg p^2$ is given in terms of sinusoidal functions, and we use these to approximate the oscillatory behavior of $\sigma$,
\bea
\sigma \approx \sigma_A \left( \frac{t}{t_0} \right)^{-3p/2}
\left[ \sin(m_\sigma t +\alpha) + \frac{-6p+9p^2}{8m_\sigma t} \cos(m_\sigma t + \alpha) \right] ~,
\label{sigmaOsci}
\eea
where $\alpha$ is a phase, and $\sigma_A$ is the initial oscillation amplitude at $t=t_0$. Such oscillations induce an oscillatory component to the Hubble parameter $H$, because
\bea
3\mpl^2 H^2 = \half \dot\sigma^2 + \half m^2\sigma^2 + {\rm other~fields}~.
\label{Heom}
\eea
The leading term on the right hand side of (\ref{Heom}) does not oscillate in time because the energy is converting between kinetic and potential energy back and forth and conserved in the leading order. The oscillatory component for $H$, which we denote as $H_{\rm osci}$, comes from the subleading terms. Using (\ref{sigmaOsci}), we get
\bea
H_{\rm osci} = - \frac{\sigma_A^2 m_\sigma}{8\mpl^2} \left(\frac{t}{t_0}\right)^{-3p} \sin(2m_\sigma t + 2\alpha) ~.
\eea
This in turn induces the oscillatory components for the parameters $\epsilon \equiv -\dot H/H^2$ and $\eta \equiv \dot\epsilon/(H\epsilon)$. Again we use the subscript ``${\rm osci}$" to denote their oscillatory components,
\bea
\epsilon_{\rm osci} &=& \frac{\sigma_A^2 m_\sigma^2}{4 \mpl^2 H^2} \left( \frac{t}{t_0} \right)^{-3p} \cos(2m_\sigma t + 2\alpha) ~,
\label{epsilon_osci}
\\
\dot\eta_{\rm osci} &=& - \frac{\sigma_A^2 m_\sigma^4}{\mpl^2 \epsilon H^3} \left( \frac{t}{t_0} \right)^{-3p} \cos(2m_\sigma t + 2\alpha)
\label{doteta_osci}
~.
\eea

\begin{figure}
\begin{center}
\epsfig{file=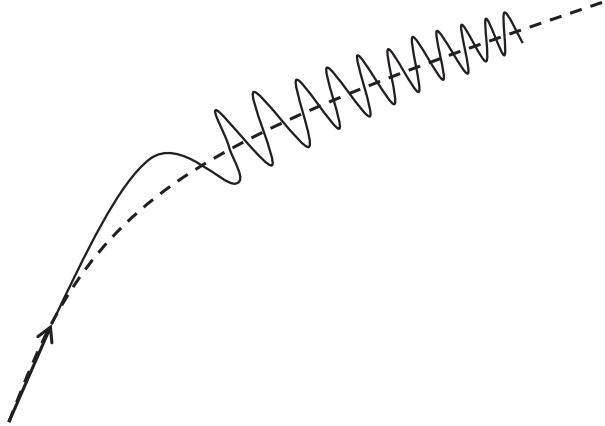, width=8cm}
\end{center}
\caption{A turning trajectory that excites the oscillation of massive fields. Dashed line indicates the potential valley. The massive field tends to settle down in the valley along the incoming and outgoing straight lines. But during the turning, the centrifugal force makes it deviate from the minimum. This induces the small oscillation.}
\label{Fig:trajectory}
\end{figure}

The next question is how these massive fields can get excited. There are many possibilities. As we have mentioned, even for single field models, we imagine a multifield configuration in which the effective single field trajectory turns from time to time depending on how the massive directions are lifted. During turning, the light mode and massive mode couple, so part of the energy can be released to excite the massive mode (Fig.~\ref{Fig:trajectory}). The resulting oscillation (\ref{sigmaOsci}) has a very high frequency. It can be averaged out in most cases, but not for our purpose. As we will see later in a more explicit example, even a tiny fraction of the energy transferred in this process can excite a large observable effect. More generally, massive fields may be excited classically by any sharp physical process, including the turning trajectory, sharp feature, particle creation and etc.

\section{Model and formalism}
\label{Sec:Model}
\setcounter{equation}{0}

We shall investigate two closely related processes and their observational signatures separately. One is the sharp feature that excites the massive fields. Another is the resonance phenomena induced by the oscillation of the excited massive fields.

We use a two-field model in the general power-law background as an example. The sharp feature happens at $t_0$. Before $t_0$, the massive particle stays at minima and we consider the single field model
\bea
\CL_1 = \sqrt{-g} \left[ -\half g^{\mu\nu} \partial_\mu \phi \partial_\nu \phi - V_\phi(\phi) \right] ~.
\label{twofield_L1}
\eea
After $t_0$, we consider the two-field model
\bea
\CL_2 = \sqrt{-g} \left[ -\half g^{\mu\nu} \partial_\mu \phi \partial_\nu \phi - V_\phi(\phi)
-\half g^{\mu\nu} \partial_\mu \sigma \partial_\nu \sigma - \half m_\sigma^2 \sigma^2 \right] ~.
\label{twofield_L2}
\eea
Around $t_0$, the $\sigma$ field is excited by some sharp process. For example, the field $\phi$ makes a turn (Fig.~\ref{Fig:trajectory}). Note that after the turn, the two fields are still decoupled if it were not for the gravity. More complicated couplings are of course possible, but this minimum case is most general. Because the $\sigma$-field is now considered as a spectator (except around $t_0$),
to study the perturbation theory, it is important that we choose the following uniform-$\phi$ gauge \cite{Chen:2009zp}, in which the scalar perturbation $\zeta$ corresponds to the conserved scalar degree of freedom in single field model,
\begin{align}
ds^2 &= -N^2 dt^2 + h_{ij}(dx^i+N^idt)(dx^j+N^jdt) ~,
\\
h_{ij} &= a^2 e^{2\zeta} \delta_{ij} ~,
\quad
\delta\phi = 0 ~,
\quad
\sigma = \sigma_0(t) + \delta\sigma(\bx,t) ~.
\end{align}
We use Maldacena's method \cite{Maldacena:2002vr} of the ADM formalism to expand the action. In this paper, we will only be interested in the correlation functions of $\zeta$, so we ignore the perturbation $\delta\sigma$. The effect of $\sigma$ comes in because its zero-mode evolution $\sigma_0(t)$ perturbs the time-dependent couplings in the perturbative expansion.

We separate the Hamiltonian in the perturbation theory as follows,
\bea
\CH_0 &=& a^3 \epsilon_0 \dot\zeta^2 - a \epsilon_0 (\partial\zeta)^2 ~,
\label{CH0}
\\
\CH_2^I &\approx& -a^3 \Delta\epsilon \dot\zeta^2 + a\Delta\epsilon (\partial\zeta)^2 + \CO(\Delta\epsilon^2) ~,
\label{CH2}
\\
\CH_3^I &\approx& -\half a^3 \epsilon \dot\eta \zeta^2\dot\zeta ~.
\label{CH3}
\eea
We have also separated $\epsilon$ into the unperturbed part and the perturbed part due to features, $\epsilon= \epsilon_0 +\Delta\epsilon$, and in $\CH_3^I$ listed the only term important for this paper. The reason that this term is important is similar to that given in \cite{Chen:2008wn,Chen:2006xjb} for inflation. Namely, the coupling in this term contains the highest time-derivative and becomes large in presence of features.

\begin{figure}
\begin{center}
\epsfig{file=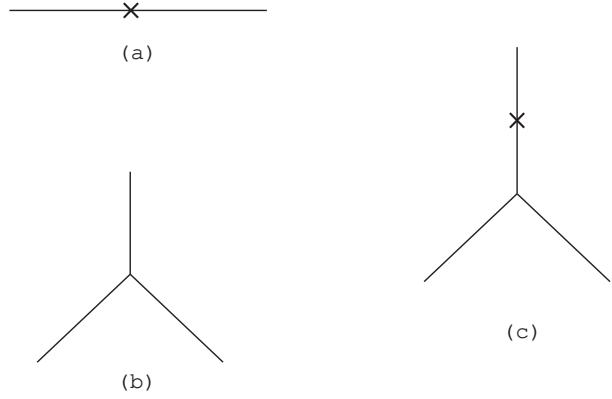, width=8cm}
\end{center}
\caption{Examples of Feynman diagrams used to perturbatively compute the power spectra and bispectra in feature models.}
\label{Fig:Fdiagram}
\end{figure}

Treating $\CH_2^I$ and $\CH_3^I$ as the interaction Hamiltonian, we can perturbatively compute the power spectrum and bispectrum using the in-in formalism,
\bea
\langle \zeta^n(t) \rangle \equiv \langle 0|
\left[ \bar T \exp \left( i\int_{-\infty}^0 a d\tau d^3x ~\CH^I \right) \right]
\zeta^n(t)
\left[ T \exp \left( -i\int_{-\infty}^0 a d\tau d^3x ~\CH^I \right) \right]
|0 \rangle ~,
\eea
where the integration of $\tau$ runs from $-\infty$ to $0$.
See \cite{Chen:2010xk} for a review of this formalism and methods for such computations.
For example, the leading correction to the power spectrum is given by the diagram Fig.~\ref{Fig:Fdiagram}(a), where the two-point vertex corresponds to (\ref{CH2}),
\bea
\Delta \langle \zeta_{\bk_1} \zeta_{\bk_2} \rangle =
\langle 0| i \int_{-\infty}^0 a d\tau d^3x ~[\CH_2^I, \zeta_{\bk_1} \zeta_{\bk_2}] |0\rangle ~.
\eea
Using the definition for the power spectrum $P_\zeta$,
\bea
\langle \zeta_{\bk_1} \zeta_{\bk_2} \rangle = \frac{P_\zeta}{2k_1^3} (2\pi)^5 \delta^3(\bk_1+\bk_2) ~,
\eea
we get
\bea
\frac{\Delta P_\zeta}{P_{\zeta 0}} = 2i \int_{-\infty}^0 d\tau~
a^2 \Delta\epsilon~
({u_{k_1}'}^2 - k_1^2 u_{k_1}^2 ) + {\rm c.c.} ~,
\label{powerspectrum_corr}
\eea
where $P_{\zeta 0}$ is the power spectrum in the absence of features, and $u_k(t)$ is the Fourier transform of $\zeta(\bx,t)$.
The leading bispectrum is given by the diagram Fig.~\ref{Fig:Fdiagram}(b), in which the three-point vertex corresponds to (\ref{CH3}),
\bea
\langle \zeta_{\bk_1} \zeta_{\bk_2} \zeta_{\bk_3} \rangle &=&
\langle 0| i \int_{-\infty}^0 a d\tau d^3x ~[\CH_3^I, \zeta_{\bk_1} \zeta_{\bk_2} \zeta_{\bk_3}] |0\rangle
\\
&=& i \left( \prod_i u_{k_i}(0) \right)
\int_{-\infty}^0 d\tau~ a^3\epsilon\dot\eta ~
u_{k_1}^* u_{k_2}^* \frac{du_{k_3}^*}{d\tau}
\nonumber \\
&\times& (2\pi)^3 \delta^3(\sum_i \bk_i) + {\rm 2~perm.} + {\rm c.c.} ~.
\label{bispectrum}
\eea
For simplicity, we will quote the bispectrum in terms of $S(k_1,k_2,k_3)$ according to the following definition \cite{Chen:2010xk},
\bea
\langle \zeta_{\bk_1} \zeta_{\bk_2} \zeta_{\bk_3} \rangle =
S(k_1,k_2,k_3) \frac{1}{(k_1k_2k_3)^2} P_{\zeta 0}^2 (2\pi)^7 \delta^3(\sum_i \bk_i) ~.
\eea

Using this method, other types of correlation functions in feature models can also be computed systematically. For example, the leading effect from the non-BD correction on the bispectrum \cite{Chen:2010bka} corresponds to the diagram Fig.~\ref{Fig:Fdiagram}(c). In this paper, we will only compute the two diagrams (\ref{powerspectrum_corr}) and (\ref{bispectrum}), which are the leading terms in this model.

In inflationary and especially non-inflationary scenarios, there are variety of ways, involving one or more fields, to produce the leading scale-invariant power spectrum $P_{\zeta 0}$ in absence of features. In this paper we do not concern how this is produced. We are interested in computing the resonance effects induced by massive fields, as corrections to the power spectrum and as the leading bispectrum.

The above formalism applies to cases with arbitrary scale factor $a(t)$. For the special case of inflation, different kinds of feature models have been studied in the past, including various effects from sharp features \cite{Starobinsky:1992ts,Wang:1999vf,Adams:2001vc,Chen:2006xjb,Hotchkiss:2009pj,Adshead:2011bw}, periodic features \cite{Chen:2008wn,Flauger:2010ja,Chen:2010bka,Leblond:2010yq}, and massive particles \cite{Burgess:2002ub,Achucarro:2010da,Jackson:2011qg}. The main point of this paper is to turn the logic around and use features to probe the background scale factor. In order to do this, it is important that we classify which type of features are observationally sensitive to different scale factors, so can be used to distinguish different paradigms; and which are not. This is also one of the issues that we will investigate in the next two sections.

\section{Sinusoidal running as trigger}
\label{Sec:Sin_trigger}
\setcounter{equation}{0}

We now study the correlation functions caused by the sharp feature around $t_0$. As emphasized, we concentrate on the universal behavior of the BD vacuum. For the kinematic Hamiltonian (\ref{CH0}), the BD vacuum behavior for the mode function
\bea
u_\bk = \int d^3x~ \zeta(t,\bx)e^{-i\bk\cdot\bx}
\eea
is the same as (\ref{BDphi}) except for different normalization factors that vary much slower than the vacuum oscillations. Namely,
\bea
u_k \to \frac{1}{a \sqrt{4\epsilon k}} e^{-ik\tau} ~.
\label{BDuk}
\eea
Although the sharp features involve both the horizon and sub-horizon scale physics, to see the most important universal feature, it is enough that we look at the subhorizon behavior (\ref{BDuk}).

Due to the sharp feature, $\epsilon$ receives some small, but sudden, change. How $\epsilon$ evolves afterwards is model-dependent. For example, in inflation, it will approach again to an attractor solution in a few Hubble time. But to see the universal effect of the sharp feature, let us only focus on this sudden change. The properties of the power spectrum and bispectrum depend on the relationship between the sharpness of this sudden change, which we denote as $\Delta\tau_s$, and the mode $k_i$.

For power spectrum, if $2k_1 \ll \Delta\tau_s^{-1}$, the feature is very sharp compared to the oscillation time-scale in BD vacuum (\ref{BDuk}). We can approximate the change in $\epsilon$ as a step function,
\bea
\Delta\epsilon \approx \epsilon_s \theta(\tau-\tau_0) ~.
\label{Deltaepsilon_appr}
\eea
Plugging (\ref{BDuk}) and (\ref{Deltaepsilon_appr}) into (\ref{powerspectrum_corr}), we get
\bea
\frac{\Delta P_\zeta}{P_{\zeta 0}} \approx \frac{\epsilon_s}{\epsilon}(1-\cos 2k_1\tau_0) ~.
\eea
The running behavior $\cos(2k_1\tau_0)$ remains similar if $2k_1 \sim \Delta\tau_s^{-1}$. For $2k_1 \gg \Delta\tau_s^{-1}$, however, the oscillation in BD vacuum is much faster and the change in $\epsilon$ is averaged out, so $\Delta P_\zeta/P_{\zeta 0} \to 0$. The important point is that, unlike the resonance case, the sinusoidal running behavior is not unique for inflation, but universal for arbitrary time-dependent background.

The case for bispectrum is similar. The sinusoidal running for $S$ is universal,
\bea
S \sim f_{NL} \cos(K \tau_0 + {\rm phase}) ~,
~~~~ K\equiv k_1+k_2+k_3 ~,
\eea
although the amplitude $f_{NL}$ now depends on the behavior of the mode function after horizon-exit, which is highly model-dependent. If we restrict to inflation, a similar estimate as in the power spectrum case can be made for the bispectrum. If $K\ll \Delta\tau_s^{-1}$, we use the approximation (\ref{Deltaepsilon_appr}) and get
\bea
f_{NL} \sim \frac{\epsilon_s}{8 \epsilon} \left( \frac{K}{Ha_0} \right)^2 ~,
\label{fNLsharp1}
\eea
where $a_0$ is the scale factor at $\tau_0$.
If $K \sim \Delta\tau_s^{-1}$, the sharpness of $\dot\eta$ and the oscillation time scale in (\ref{BDuk}) are comparable, so (\ref{bispectrum}) leads to
\bea
f_{NL} \sim \frac{\Delta \eta}{\Delta\tau_s a_0 H}
\sim \frac{\epsilon_s}{\epsilon} \frac{1}{(H a_0 \Delta\tau_s)^2} ~.
\label{fNLsharp2}
\eea
If $K \gg \Delta\tau_s^{-1}$, $f_{NL} \to 0$.

So the correction to the power spectrum is generally very small.
For example, specializing the model of Sec.~\ref{Sec:Model} to the slow-roll inflation case, we have $\epsilon_s/\epsilon \sim \beta$, where $\beta$ is the fraction of the kinetic energy of $\phi$ converted to that of $\sigma$.
The size of the bispectrum (\ref{fNLsharp2}) depends on $\Delta\tau_s$ and can be very large if the feature is sharp ($H a_0 \Delta\tau_s = H \Delta t_s \to 0$).

To summarize, first, the most important property caused by sharp feature is the sinusoidal running for modes $K\gg \tau_0^{-1}$; this shows up as the correction to the power spectrum,
\bea
\frac{\Delta P_\zeta}{P_{\zeta 0}} \propto \sin (2k_1\tau_0 + {\rm phase}) ~,
\eea
and as the leading contribution in the bispectrum,
\bea
S \propto \sin (K \tau_0 + {\rm phase}) ~.
\eea
Second, the starting point for this running is around the scale $k_0\equiv |\tau_0|^{-1}$, which is the mode that is crossing the event-horizon at the time of the feature $t_0$; the wavelength of this sinusoidal running in $2k_1$-(or $K$-)space is given by the same scale $2\pi k_0$. Third, this qualitative behavior is universal for arbitrary time-dependent background (i.e.~for all values of $p$);\footnote{For inflation, this type of running has been shown for power spectra \cite{Starobinsky:1992ts,Adams:2001vc,Achucarro:2010da} and bispectra \cite{Chen:2006xjb,Chen:2008wn,Hotchkiss:2009pj,Adshead:2011bw}.
For cases where the feature is very sharp so modes well within the horizon can be affected, the formulae (\ref{fNLsharp2}) gives a better estimation for the maximum bispectrum amplitude than those in \cite{Chen:2006xjb,Chen:2010xk}. For example, for a small step in slow-roll potential with width $d$ and relative height $c$, $f_{NL}^{\rm max} \sim c(c+\epsilon)/(d^2\epsilon)$.} they cannot be used to distinguish inflation from the alternatives, but can be used to identify the location of the sharp feature, which is a signal that some massive fields are likely to be excited. In the next section, we will study the effects of these massive fields on density perturbations, including their profiles, locations and magnitudes. These will become the main signals we use to distinguish different primordial universe paradigms.

We briefly comment that there may be other model-dependent signatures due to interaction with massive modes. For example, if we consider the two-field model in Sec.~\ref{Sec:Model} in inflationary spacetime, during the sharp turn, quantum fluctuations of the massive modes are projected to the curvature mode. By matching the curvaton mode function before and after $\tau_0$, we can see that the factional correction to the power spectrum due to this effect is $\Delta P_\zeta/P_{\zeta 0} \sim \theta_0 (m/H) (-k_1\tau_0)^{3/2} \cos[(m/H)\ln k_1 + {\rm phase}(k_1)]$, where $k_1<\tau_0^{-1}$ and $\theta_0$ is the turning angle. The ${\rm phase}(k_1)$ is a $k_1$-dependent random phase from the massive modes, due to which there is no definite prediction on the oscillatory running. But the main signature is the overall amplitude with a blue tilt, $\sim k_1^{3/2}$, because massive fluctuations decay in expanding spacetime. These predictions are more model-dependent and we will not discuss them in more details in this paper. They may provide supportive evidence on the detailed process.

\section{Resonant running as evidence}
\setcounter{equation}{0}

We now compute the resonance effect on power spectrum and bispectrum induced by the excited oscillatory massive fields. As in the previous section, we first compute the correlation functions in the general power-low background, and point out the most significant general behavior. When we wish to see whether such effects are large enough to be observable, we use inflation as the explicit example. This is because our main purpose here is to find distinctive signatures for inflation. Otherwise, one can use concrete alternative models as explicit examples.

All the necessary ingredients for this computation are ready. The same formalism in Sec.~\ref{Sec:Model} applies here. For power spectrum, we use (\ref{epsilon_osci}) for $\Delta\epsilon$ in (\ref{powerspectrum_corr}). From Sec.~\ref{Sec:BDandres}, we know that for resonance we only need the universal BD behavior (\ref{BDuk}) for $u_k$ in (\ref{powerspectrum_corr}). After performing the same type of integral encountered in Sec.~\ref{Sec:BDandres}, we get
\bea
\frac{\Delta P_\zeta}{P_{\zeta 0}} =
\frac{\sqrt{\pi}}{4} \frac{\sigma_A^2}{\epsilon \mpl^2}
\left( \frac{m_\sigma}{H_0} \right)^{5/2}
\left( \frac{2k_1}{k_r} \right)^{-3+\frac{5}{2p}}
\sin \left[ \frac{p^2}{1-p} \frac{2m_\sigma}{H_0} \left( \frac{2k_1}{k_r} \right)^{1/p} -2\alpha + \frac{3\pi}{4} \right] ~,
\label{res_powerspectrum}
\eea
where $H_0$ are evaluated at $t_0$, and, for power-law scale factor, $\epsilon=1/p$ is constant.
For bispectrum, we use (\ref{doteta_osci}) for $\dot\eta$ in (\ref{bispectrum}). The general amplitude is model-dependent, but the resonant running behavior is given by
\bea
S \propto \left( \frac{K}{k_r} \right)^{-3+\frac{7}{2p}}
\sin \left[ \frac{p^2}{1-p} \frac{2m_\sigma}{H_0}
\left( \frac{K}{k_r} \right)^{1/p} + {\rm phase} \right] ~,
\label{res_bispectrum}
\eea
where we have also included a $K$-dependent modulation factor which typically arises but is not as robust as the rest of the running behavior.

In these results, we have defined a parameter $k_r$ which denotes the first $K$-mode that resonates as soon as the massive field starts to oscillate at $t_0$. Namely, $k_r \equiv 2m_\sigma a_0$. Recall that, at $t_0$, the comoving mass-scale of the event horizon is $k_0 \equiv |\tau_0|^{-1}$; and $k_0$ is the starting mode in the sinusoidal running due to sharp feature. It is important to notice that there is a relation between the ratio $k_r/k_0$ and the ratio $2m_\sigma/H_0$,
\bea
\frac{k_r}{k_0} = \frac{|p|}{|1-p|} \frac{2m_\sigma}{H_0} ~.
\eea

We can also qualitatively understand how the resonant running is capable of recording the scale factor evolution. The oscillating massive field provides periodically oscillating background, as well as a resonance scale with constant physical wave-number. Different $K$-modes of the BD vacuum are stretched or contracted by the scale factor $a(t)$, and resonate with the background when their physical wave-number coincide with the resonance scale. When the change in $K$ corresponds to the change in $t$ that is equal to the oscillation period of the massive mode, the final phase grows by $2\pi$. This is why the arguments of the sinusoidal functions in (\ref{res_powerspectrum}) and (\ref{res_bispectrum}) are power-law function with the inverse power $1/p$, which is the inverse function of the power-law in scale factor.

Take the exponential inflation limit, $p\gg 1$, in the two-field model.\footnote{For inflation, the effect of the resonance mechanism on power spectrum and non-Gaussianity due to periodic features in single field models is studied in \cite{Chen:2008wn,Flauger:2010ja,Chen:2010bka,Leblond:2010yq}; the effect of oscillating massive field at the beginning of inflation on power spectrum is studied in \cite{Burgess:2002ub} by introducing a direct coupling to inflaton; the effect on power spectrum after integrating out the massive modes is studied in \cite{Jackson:2011qg}.} For power spectrum, we get
\bea
\frac{\Delta P_\zeta}{P_{\zeta 0}} =
\frac{\sqrt{\pi}}{4} \frac{\sigma_A^2}{\epsilon \mpl^2}
\left( \frac{m_\sigma}{H} \right)^{5/2}
\left( \frac{2k_1}{k_r} \right)^{-3}
\sin \left[ \frac{2m_\sigma}{H} \ln 2k_1 + \tilde \alpha \right] ~,
\label{Res_2pt_inf}
\eea
where the phase $\tilde \alpha = (2m_\sigma/H)(1-\ln 2m_\sigma) +2\alpha+\pi/4$.
For bispectrum,
\bea
S = \frac{\sqrt{\pi}}{8}
\frac{\sigma_A^2}{\epsilon \mpl^2}
\left( \frac{m_\sigma}{H} \right)^{9/2}
\left( \frac{K}{k_r} \right)^{-3}
\sin \left[ \frac{2m_\sigma}{H} \ln K + \hat\alpha \right] ~,
\label{Res_3pt_inf}
\eea
where $\hat\alpha = (2m_\sigma/H)(1-\ln 2m_\sigma) + 2\alpha- 3\pi/4$.
Both (\ref{Res_2pt_inf}) and (\ref{Res_3pt_inf}) take the CEL form (\ref{CELform}).

\begin{figure}[htpb]
\begin{center}
\epsfig{file=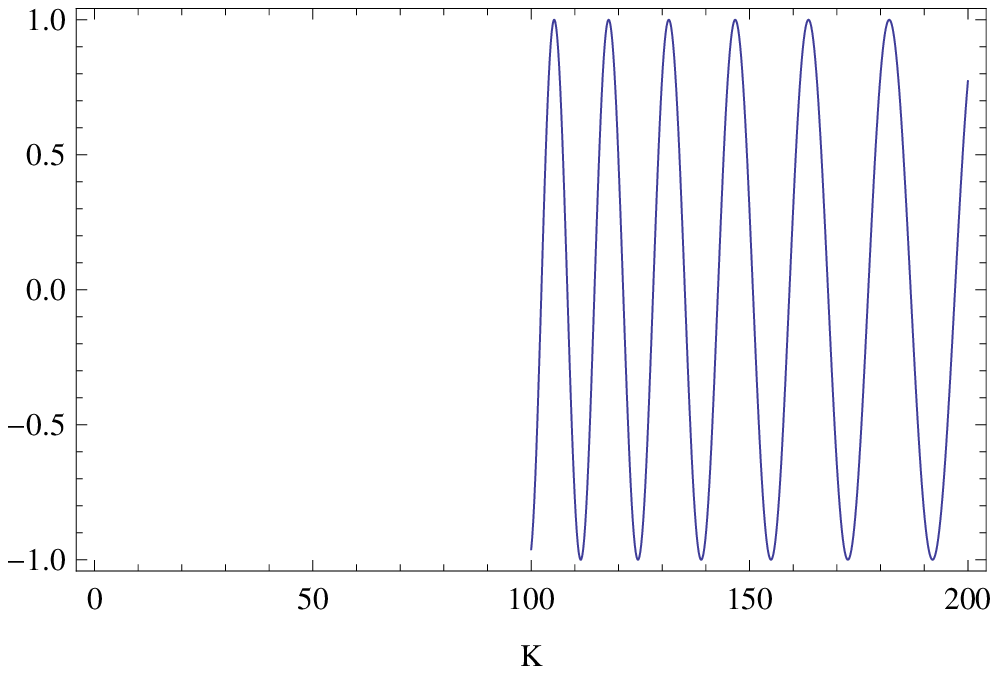, width=0.5\textwidth}
\epsfig{file=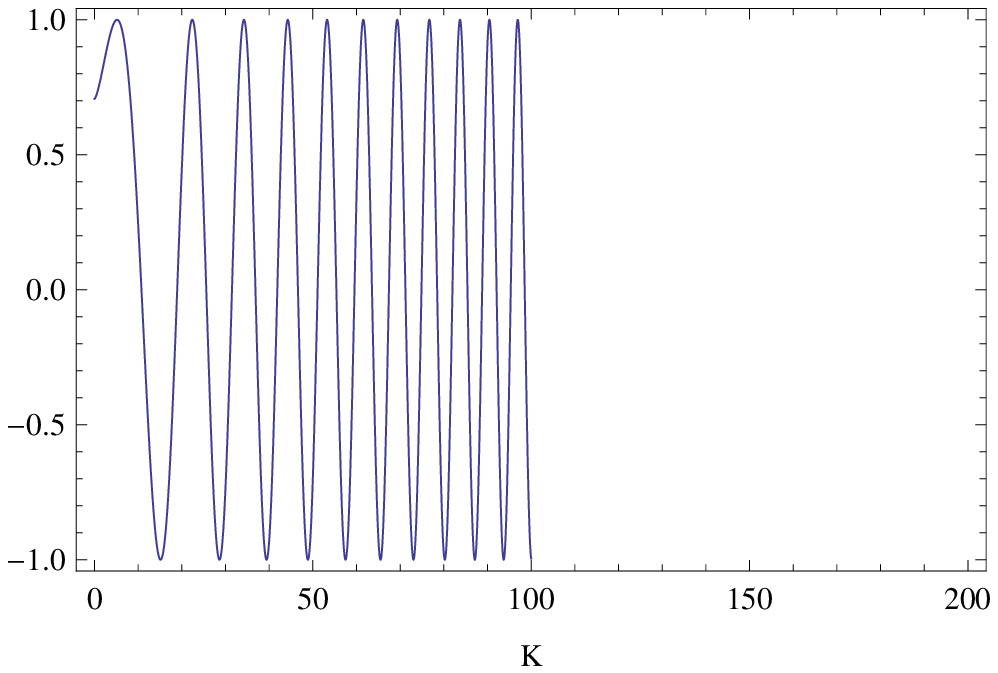, width=0.5\textwidth}
\epsfig{file=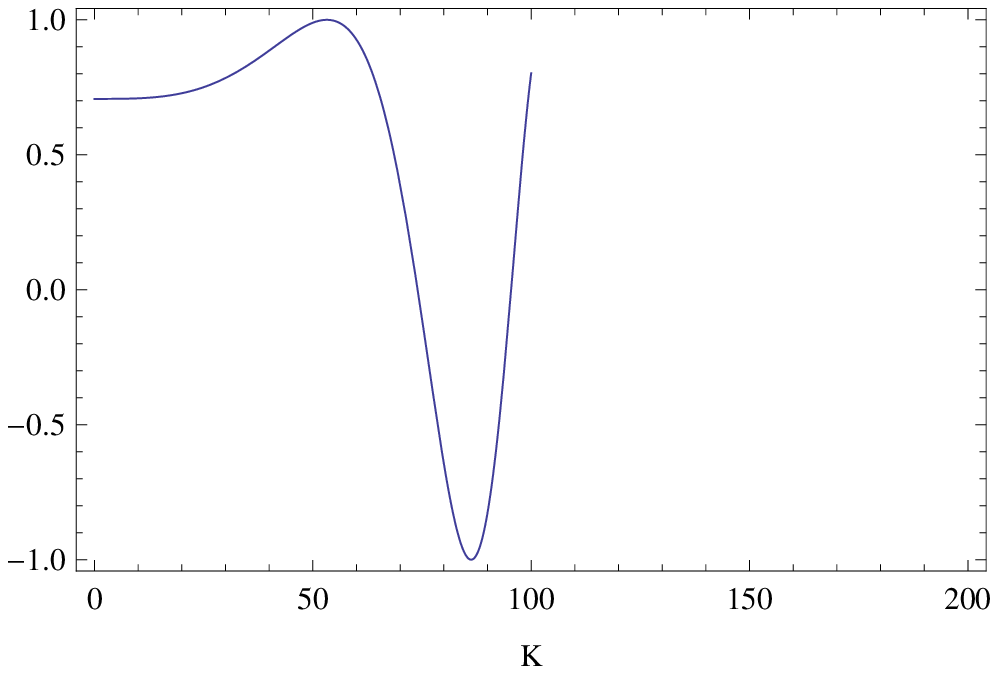, width=0.5\textwidth}
\end{center}
\caption{Resonance running in different time-dependent backgrounds due to features periodic in time, $\sin\left[ \frac{p^2}{1-p} C (K/k_r)^{1/p} + {\rm phase} \right]$. Note that this does not include the running of the amplitudes. In these plots we use $C=2m_\sigma/H_0=50$, $k_r=100$, ${\rm phase} =\pi/4$; and from top to bottom, $p=10$ (inflation), $2/3$ (matter contraction), $0.3$ (Ekpyrosis).}
\label{Fig:res_running}
\end{figure}

\subsection{Resonant running}

The resonant running for different $p$ are very different. Let us look at the details in the power spectra. The bispectra are same after replacing $2k_1$ with $K$.

First, $k_r$ is the first resonant mode, but for different $p$ the subsequent resonant modes are different. For the expanding background $p>1$ and $p<0$, both (\ref{res_powerspectrum}) and (\ref{res_bispectrum}) apply only for $2k_1>k_r$, since lower $k$-modes resonate earlier.  For the contracting background $0<p<1$, the situation is opposite and the results apply only for $2k_1<k_r$.

Second, if we denote the local periodicity of the resonant running in $k$-space as $\Delta k_1$, we have $\Delta k_1 \propto k_1^{-1/p+1}$. So for $p>1$ and $p<0$, $\Delta k_1$ increases as $k_1$ increases; while for $0<p<1$, $\Delta k_1$ increases as $k_1$ decreases.

Several examples of resonant running are plotted in Fig.~\ref{Fig:res_running}.
As emphasized, the differences in these resonant runnings are kept intact even after general curvaton-isocurvaton transformation, and they are the faithful signals we can use to distinguish the primordial universe paradigms.

\subsection{Running of amplitudes}

Besides the difference in the resonance running, different spacetime backgrounds also give rise to different scale-dependence in the modulation amplitudes.
These scale dependence are much milder because the range of scales $\Delta K$ over which the variation takes place is larger than the local scale $K$ itself. For more complicated multi-field models, such scale dependence can be changed due to scale-dependence in the curvaton-isocurvaton couplings. So in general they are not always the faithful signatures for our purpose. However, there are some dramatic properties which may be useful as supportive evidences, so let us nonetheless examine these properties. We define the running indices
\bea
n_p \equiv \frac{d \ln (\Delta P_\zeta/P_{\zeta 0})_A}{d \ln k}
\eea
for power spectrum, and
\bea
n_b \equiv \frac{d\ln f_{NL}}{d \ln K}
\eea
for bispectrum. The $(\Delta P_\zeta/P_{\zeta 0})_A$ and $f_{NL}$ denote the modulation amplitudes, i.e.~the overall factors in front of the sinusoidal functions, in $\Delta P_\zeta/P_{\zeta 0}$ and $S$, respectively.\footnote{Note that for bispectra we have separated the resonant running from the definition of $f_{NL}$, as in \cite{Chen:2010xk}. So the index is slightly different from the definition $n_{NG}-1$ in \cite{Chen:2005fe,Byrnes:2009pe}, where it is defined for bispectra with non-oscillatory running. } So for power spectrum we have
\bea
n_p= -3 + \frac{5}{2p} ~,
\label{index_np}
\eea
and for bispectrum we typically have
\bea
n_b = -3 + \frac{7}{2p} ~.
\label{index_nb}
\eea

For exponential inflation, $p\gg 1$, both indices are red, $n_p=n_b=-3$. The factor $-3$ is present for all expanding backgrounds because the amplitude of the massive mode is damped by the expansion as $t^{-3p/2}$. This factor is also present for the contracting backgrounds for the following reason. For contracting backgrounds ($0<p<1$), the amplitude of massive mode is growing as $t^{-3p/2}$ (recall $t$ runs from $-\infty$ to $0$ in this case). But an important difference between the expanding and contracting background is that, in the former, smaller $k$-modes resonate earlier, but in the latter, larger $k$-modes resonate earlier. This is why although the amplitude of the massive mode evolves oppositely in time for the two cases, their contribution to the running index turns out to be the same.

There is also an additional factor $\sim 1/p$. For contracting background with small $p$, this makes both indices blue, $n_p \approx 5/(2p)$ and $n_b \approx 7/(2p)$. This factor is due to two reasons. First, for contracting background, the resonance scale is fixed while the event horizon is shrinking. So the resonance strength ($\propto |t|^{1/2}$) gets weaker for smaller $k$-modes. In the meanwhile, the couplings (\ref{epsilon_osci}) and (\ref{doteta_osci}) depend on $H$, which ($\propto |t|^2$ and $|t|^3$ respectively) also get weaker for smaller $k$.

As discussed, the details of the indices may not be faithfully kept in terms of the curvature mode in more general models; but the dramatic difference between the different cases, such as $p\gg 1$ and $0<p\ll 1$, can serve as supportive evidence.

We also comment that, because the blue or red running indices are generally of order one or larger, the signals we are looking for typically decay away in a few efolds. But this does not limit their usage. As we can see from the last subsection and Fig.~\ref{Fig:res_running}, the differences in resonant running for different $p$ are already very clear within a couple of efolds.

\subsection{Amplitudes}

We use the two-field model (\ref{twofield_L1}) and (\ref{twofield_L2}) to show that the amplitudes of the resonant power spectra and bispectra can be easily made very large, at least for the inflation models.

Consider the example of slow-roll inflation. We denote the fraction of the kinetic energy of $\phi$, that is converted to the energy in the $\sigma$-field during the turning and induces its oscillation, as $\beta$. So $m_\sigma^2 \sigma_A^2 \sim \beta \dot\phi^2$. Also note $\epsilon \sim \dot\phi^2/(\mpl^2 H^2)$. From (\ref{Res_2pt_inf}) and (\ref{Res_3pt_inf}), we have
\bea
\left( \frac{\Delta P_\zeta}{P_{\zeta 0}} \right)_A
&\sim& \beta \left( \frac{m_\sigma}{H} \right)^{1/2} ~,
\label{Est_Power}
\\
f_{NL} &\sim& \beta \left( \frac{m_\sigma}{H} \right)^{5/2} ~.
\label{Est_Bispectrum}
\eea
So even for a tiny fraction of energy transfer, the resonance amplitudes can be quite large. For example, for $\beta \sim 10^{-2}$, $m_\sigma/H \sim 10^2$, we have $\Delta P_\zeta/P_{\zeta 0} \sim 0.1$ and $f_{NL} \sim 10^3$.
Because of the spatial inhomogeneity characterized by $\delta t \sim 10^{-5}/H$, the zero-mode oscillation in the classical background receives a random phase correction, $\omega \delta t$. This phase has to be much smaller than $2\pi$ so that the signals we are interested are not averaged away. Therefore we can at most explore the massive modes over five order of magnitudes above $H$, $m_\sigma/H < 10^5$.

\section{The signature pattern for inflation}
\label{Sec:Signature}
\setcounter{equation}{0}

We provide a summary on the signature pattern for the inflation paradigm. Similar summary can also be done for each alternative paradigm by specializing the previous general results.

We have shown that a detection of the resonant form of CEL type induced by massive field oscillation in power spectrum or non-Gaussianities is an evidence for the inflation paradigm. But to be unambiguous, it is important to strengthen the evidence that this is due to the periodically oscillating massive fields, by using other characteristic properties besides the resonant running. The following are the signature pattern that we can look for in density perturbations:

\begin{itemize}

\item {\em Trigger.} Observational signatures associated with sharp feature can be used as a sign that some massive fields may be excited. These signatures appear as sinusoidal running in density perturbations, as corrections to power spectrum or dominant components in non-Gaussianities. These oscillations start at a scale $k_0$, have constant wave-length $\sim 2\pi k_0$ in $2k_1$-(or $K$-)space, and propagate towards larger $k$-modes with model-dependent growing and then decaying amplitudes.

\item {\em Signal.} Excited massive field induces highly oscillatory resonant running in density perturbations, as corrections to power spectrum or dominant components in non-Gaussianities. This resonant running has a distinct CEL form (\ref{CELform}). For example for bispectra, it starts at a scale $k_r$ and propagate towards larger $K$-modes, typically with decaying amplitude and lasting for no more than a few efolds. The oscillating wavelength $\Delta K$ is always smaller than the local $K$, with a fixed ratio that is determined by the parameter $2m_\sigma/H$, i.e.~$\Delta K/K = \pi H/m_\sigma$. The starting place $k_0$ for the previous sinusoidal running and $k_r$ for this resonant running is related by the same parameter $2m_\sigma/H$, i.e.~$k_r/k_0 = 2m_\sigma/H$.

    It is also likely that several massive fields with different mass are excited at the same time. So we may look for different CEL forms in modes much larger $k_0$, each satisfying the relation $k_{r i}/k_0 = 2m_{\sigma i}/H$. The relation between the different CEL forms can also be used to conclude that they are induced by the same sharp feature, even in case where the observational signatures from the sharp feature is too weak to be observable. Namely, by measuring $k_r$ and $2m_\sigma/H$ for each form, they should satisfy
    \bea
    \frac{k_{r1}}{2m_{\sigma 1}/H}= \frac{k_{r2}}{2m_{\sigma 2}/H} = \cdots ~.
    \eea

\item {\em Caveats and solutions.} It is also important to note several caveats and possible solutions.

    In the inflation case we considered above, the mass may be time-dependent. But such dependence has to be very dramatic [$\dot m_\sigma/(m_\sigma H) \gtrsim \CO(1)$] to make the final resonance form differ significantly from the CEL form.\footnote{The dramatic time-dependence in mass will also lead to large running in the oscillating amplitudes (\ref{Est_Power}) and (\ref{Est_Bispectrum}), therefore modifying the overall running behavior of the amplitudes.} So for inflation with massive modes, the CEL form is the generic form we expect to measure. The question we concern is how non-inflationary spacetime may produce the same specific pattern.

    For non-inflationary spacetime, periodic oscillations from massive modes generate different types of resonant forms (\ref{ResPowerlaw}). So to mimic the CEL form we need to engineer artificial features. As we have shown in Sec.~\ref{Sec:BDandres}, features that introduce a background oscillation component of the form $B(t) \sim e^{ig \ln(t/t_0)}$ can also induce the CEL form. To reproduce the signature pattern for inflation, we need to place a sharp feature right at the beginning of these repeated features, to satisfy the relation for $k_r/k_0$. A possible solution to such an ambiguity is to detect or constrain more observables, which naturally arise in the same inflation model, but makes reverse engineering in the alternatives more artificial. For example, as we mentioned, it is natural that more than one set of CEL forms are present with different $2m_\sigma/H$. To engineer them in non-inflationary spacetime, we need to superimpose repeated features with different $g$ parameter on top of each other, and right after the sharp feature. In addition, the characteristic running amplitudes of the resonance forms, (\ref{index_np}) and (\ref{index_nb}), can be used as supportive evidence. Furthermore, a sharp feature in non-inflationary case is likely to excite massive fields, which induce different types of resonant forms. Constraining these forms can provide additional supportive evidence.

    Finally, although we expect small excitations of some massive fields exist generically, they are not always observable. For example, for density perturbations at $\ell \sim \CO(10^3)$, the highest mass we can possibly detect through this method is $\CO(10^3) H$. This is most likely to be further limited by experimental sensitivities and sky coverage. Therefore experiments that target on high multipoles are most useful for our purpose.

    Overall, like the tensor modes, resonance phenomenon induced by massive fields has its generic and distinctive set of predictions for general inflation models.
    In addition, the signatures for different paradigms are different and can be used to distinguish inflation from the other paradigms without degeneracy; this aspect is even more advantageous than the tensor modes.
    But it also has similar caveats. Not all parameter space are measurable. Also they may be engineered in alternative paradigms by different processes, but such engineering can become highly artificial by predicting, constraining and measuring more observables naturally present in such phenomena.

\end{itemize}

\section{Experiments and data analyses}
\setcounter{equation}{0}

As we know, the tensor mode is determined by the horizon mass-scale in the primordial universe, such as the Hubble parameter $H$ in inflation, which may be much higher than energy scales accessible in accelerators. Interestingly the mechanisms studied here involve energies much larger than $H$, and therefore is a probe of even higher energy scales. In this paper we have shown that such mechanisms can record the time-dependence of the scale factor of the primordial universe in terms of distinctive oscillatory running of resonance forms. Such features are determined by the properties of the BD vacuum that is shared by all scenarios, and are kept intact for general multifield models.

In this section, we discuss several experimental and data analyses aspects.
As indicated by the CEL form, such effects show up in terms of oscillatory signals in $k$-space. To observe them, the binning in the multipole space $\Delta \ell$ has to be much smaller than $\ell$ itself. This requires high precision experiments capable of observing density perturbations for large $\ell$. The Planck satellite is observing the CMB at maximum multipoles of a few thousands. The ground-based telescopes, Atacama Cosmology Telescope (ACT) \cite{ACT} and South Pole Telescope (SPT) \cite{SPT}, can go up to ten thousands. More speculatively, the 21cm hydrogen line may be observed in much lower redshift and in much higher multipoles.

The CEL form, and most other resonance forms, have highly oscillatory and characteristic running behavior. This makes it very difficult for other effects, such as the astrophysical, nonlinear gravity and systematic effects, to mimic such signals.
For example for CMB, the observational sensitivities for conventional power spectrum and bispectra are dramatically reduced as we go to high $\ell$ of several thousands, due to astrophysical effects such as the point sources and the Sunyaev-Zeldovich (SZ) effect. This is the case for the ACT and SPT experiments in the range $\ell \gtrsim 2500$.
Since the CEL form is orthogonal to these contaminations, part of this range may now become important in terms of probing the primordial cosmology.
For CMB experiments, the nonlinear effects in CMB evolution, which limit the sensitivity for various scale-invariant bispectra to be of order $f_{NL} \sim \CO(1)$, are also orthogonal to the type of signals we study here. Therefore a new assessment is necessary to find out the main limiting factors and make forecasts.

To search for such signals in CMB, instead of starting with a specific template, we need to scan a variety of non-separable functional forms. The modal decomposition method \cite{Fergusson:2006pr,Fergusson:2008ra} developed by Fergusson, Shellard and collaborators seems ideal for such goals. This method has been mainly applied to general bispectra \cite{Fergusson:2009nv,Fergusson:2010dm} and trispectra \cite{Regan:2010cn,Fergusson:2010gn} with less dramatic scale dependence, but should be able to be generalized to cases with highly oscillatory scale dependence, as well as to the power spectrum.

\medskip
\section*{Acknowledgments}
I would like to thank Niayesh Afshordi, James Fergusson, Eugene Lim, Paul Shellard and Meng Su for helpful discussions. I am supported by the Stephen Hawking advanced fellowship.

\end{spacing}

\end{document}